\documentclass[prd, preprint, 12pt]{revtex4-1}

\usepackage{amsmath}
\usepackage{amssymb}
\usepackage{setspace}
\usepackage{graphicx}
\usepackage{natbib}
\usepackage{float}
\usepackage{tensor}
\usepackage[utf8]{inputenc}
\usepackage{amsfonts}
\usepackage{fdsymbol}
\usepackage{braket}

\begin{document}
 
 % ! TEX spellcheck
 %
%\ \vskip 1.0 in

\begin{center}
 { \large {\bf Quantum Gravity as an Emergent Phenomenon}}

%\smallskip

\vskip 0.3 in

{\large{\bf Shounak De$^{\&}$, Tejinder P.  Singh$^*$ and Abhinav Varma$^\#$}}

%{\it $^{*}$Indian Institute of Technology Bombay, Powai, Mumbai 400076, India}\\  
{\it Tata Institute of Fundamental Research,}
{\it Homi Bhabha Road, Mumbai 400005, India}\\
\bigskip
{$^\&$\tt shounakde@alumni.iitg.ac.in},
{$^*$\tt tpsingh@tifr.res.in} [Corresponding Author], 
{$^\#$\tt abhinavvarma1995@yahoo.co.in}

\end{center}

\bigskip
\bigskip

\centerline{\bf ABSTRACT}
\noindent There ought to exist a reformulation of quantum theory which does not depend on classical time. To achieve such a reformulation, we introduce the concept of an atom of space-time-matter (STM). An STM atom is a 
classical non-commutative geometry based on an asymmetric metric, which is sourced by a closed string. Different  atoms interact via entanglement. The statistical thermodynamics of a large number of such atoms gives rise to
a theory of quantum gravity, at equilibrium. Far from equilibrium, where statistical fluctuations are large, the emergent theory reduces to classical general relativity. In this theory, classical  black holes are far from equilibrium low entropy states, and their Hawking evaporation represents an attempt to return to the [maximum entropy] equilibrium quantum gravitational state.
\bigskip
\noindent 

\noindent 

\vskip 1 in

\centerline{March 31, 2019}
%submitted to GRF March 31

\bigskip

\centerline{Essay written for the Gravity Research Foundation 2019 Awards for Essays on Gravitation}
\centerline{Honorable Mention, to appear in {\it Int. J. Mod. Phys.}}
\bigskip

%\centerline {{\bf Corresponding Author:} Tejinder P. Singh}
\bigskip
%\centerline{\it This essay received an honorable mention in the Gravity Research Foundation 2016 Essay Contest}

\newpage

\setstretch{1.25}
\noindent It is a consequence of the Einstein hole argument that there must exist a reformulation of quantum theory which does not refer to a classical space-time manifold \cite{Carlip2001}. The search for such a reformulation is the most pressing amongst all the foundational problems of quantum theory, and such a search also naturally leads us to a candidate  quantum theory of gravity \cite{Singh2019qg}. Considering that non-commutativity in phase space is the essence of quantum theory, the natural substitute for space-time in the sought for reformulation is the algebra of non-commuting coordinate operators in Connes' non-commutative geometry. We propose here that physical laws are invariant under general coordinate transformations of non-commuting coordinates. This helps us to construct a non-commutative classical gravity theory, from which there emerges the sought for reformulation of quantum theory without classical time.

When coordinates are non-commuting, anti-symmetric objects such as $(d\hat{t}\; d\hat{x} - d\hat{x}\; d \hat{t})$ are non-vanishing, and contribute to the line element. Hence we build the non-commutative gravity theory by starting from a classical (commuting) gravity-torsion theory \cite{Hammond} based on an asymmetric metric $\Phi_{\mu\nu} \equiv g_{\mu\nu} + \psi_{\mu\nu}$ 
and the asymmetric connection 
\begin{equation}
\tilde{\Gamma} \indices{_{\mu \nu}^{\sigma}} = \Gamma \indices{_{\mu \nu}^{\sigma}} + S \indices{_{\mu \nu}^{\sigma}} \
\label{3a}
\end{equation}
The symmetric part of the connection (Christoffel symbols)  is related to the symmetric part of the metric as usual, and we define the anti-symmetric part of the connection (i.e. the torsion tensor, which in this case is completely  anti-symmetric) from the  anti-symmetric part of the metric:
\begin{align}
S \indices{_{\mu \nu \sigma}} = \psi \indices{_{[\mu \nu, \sigma]}} = \frac{1}{3} (\psi \indices{_{\mu \nu, \sigma}} + \psi \indices{_{\sigma \mu, \nu}} + \psi \indices{_{\nu \sigma, \mu}})
\label{2a}
\end{align}
The gravity-torsion part of the action is the (dimensionless) Einstein-Hilbert action for an asymmetric metric and asymmetric  connection
\begin{align}
\mathcal{S}_g = \frac{1}{L_p^2} \int d^4x \;\sqrt{-\Phi} \; R(\tilde{\Gamma}) 
\label{2b}
\end{align}
The only fundamental constants are the square of Planck length, and the speed of light. Newton's gravitational constant and Planck's constant $\hbar$ will be emergent subsequently.
It can be shown that the vacuum field equations arising from this action do not admit spherically symmetric solutions, and that the matter source must be an extended object which possesses a spin vector $\xi^\mu$, apart from mass:
\begin{align}
\mathcal{S}_M = \sum_{n} \bigg[ m_n c^2 \int d \tau_n \, \sqrt{\frac{dx^{\mu}_{n}}{d\tau_n}\frac{dx^{\nu}_{n}}{d\tau_n} g_{\mu \nu}} + \frac{c^2}{2} \int d\tau_n \, \xi_{n}^{\mu} \frac{dx^{\nu}_{n}}{d\tau_n} (2 \psi_{\mu \nu} + g_{\mu \nu}) \bigg] 
\label{6a}
\end{align}
\newpage
\begin{figure}[H]
	\centering
	\includegraphics[width=1.1\linewidth]{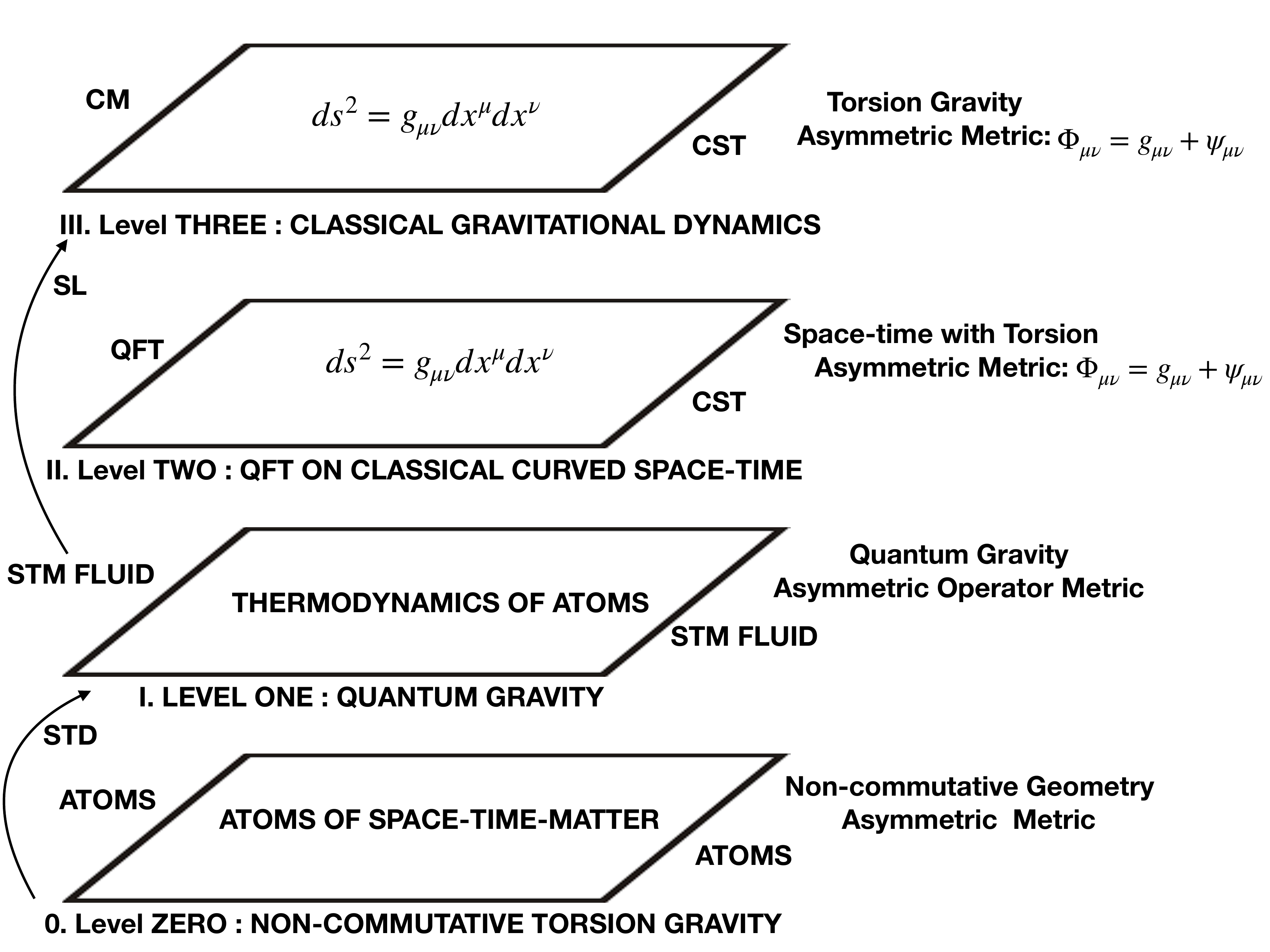}
	\caption{The four layers of gravitational dynamics. {\bf Level III} is classical gravity: a torsion gravity theory with an asymmetric metric, coupled to strings, to which classical general relativity is an 
	excellent approximation in the macroscopic realm. (CM: Classical Mechanics, CST: Classical Space-Time). {\bf Level 0}: In the microscopic realm, torsion becomes significant, but one then has to introduce non-commutativity in 
	the space-time geometry. The theory at Level 0 is non-commutative torsion gravity - a classical theory. Each space-time-matter (STM) atom is a non-commutative torsion gravity; STM atoms interact with each other through entanglement. {\bf Level I} is quantum field theory without classical time,  and it is also quantum gravity. It is arrived at from Level 0 by doing the statistical thermodynamics (STD) of the STM atoms of Level 0. The transition from  Level I to Level III is via spontaneous localisation (SL). This causes the emergence of classical space-time and classical general relativity, while also explaining the quantum-to-classical transition and providing a falsifiable solution to the quantum measurement problem.  {\bf Level II}: Quantum field theory on a classical
	space-time, is a hybrid of Level I and Level III: matter is quantum but space-time geometry is classical. We explain in the essay the conditions under which such a hybrid description is valid. Figure taken from \cite{Singh2019qg}.}
\end{figure}

The spin couples to torsion, as expected. For macroscopic sources, mass dominates spin, torsion is negligible, and the theory reduces to general relativity. Remarkably, this matter action above is the low energy limit of the world-sheet action for a closed string \cite{Hammond}
\begin{align}
\mathcal{S}_{M} = \frac{1}{L_1^2} \int \sqrt{- \gamma}\; d^2 \chi + \frac{1}{L_2^2} \int \psi_{\mu \nu}\; d \sigma^{\mu \nu} \,,
\label{10a}
\end{align}
which includes a Nambu-Goto part and a Kalb-Ramond part. We call the theory described by the action $S\equiv S_g+S_M$
\begin{align}
\mathcal{S} =  \frac{1}{L_p^2} \int d^4x \sqrt{-\Phi} \, R(\tilde{\Gamma}) + \frac{1}{L_1^2} \int \sqrt{- \gamma}\; d^2 \chi + \frac{1}{L_2^2} \int \psi_{\mu \nu}\; d \sigma^{\mu \nu} \,,
\label{fullacn}
\end{align}
(commutative) torsion-gravity. This is the theory at Level III in our Level diagram in Fig. 1. Its field equations can be derived by varying w.r.t. the metric, and the string evolves according to a geodesic equation which follows from the field equations.

Our fundamental theory at Level 0, from which quantum theory without classical time will emerge, is the non-commutative analog of the above (commutative) torsion-gravity theory. In Connes' 
non-commutative geometry (NCG), the role of conventional real coordinates $\{x^\mu\}$ is taken over by operators  $\hat{x}^\mu$ in the Hilbert space ${\cal H}$. The infinitesimal $dx$ is replaced by a compact operator $\epsilon$, the integral  $\intbar \epsilon$ of an infinitesimal  is the coefficient of the logarithmic divergence in the trace of $\epsilon$. 
The Riemannean line-element $ds = \sqrt{g_{\mu\nu}dx^\mu dx^\nu}$ is replaced by the operator $d\hat s$ = fermion propagator, i.e. inverse of the Dirac operator $D$ \cite{Connes2000}.

The algebra ${\cal A}$ of the operator coordinates, along with the Hilbert space and the Dirac operator, define the spectral triple $\{{\cal A}, {\cal H}, D\}$ which is the essence of NCG. 
In NCG, we no longer have the Riemannian manifold of space-time as a fundamental entity, but rather, it is an approximate  derived concept, constructed out of the $D^{-1}$ which smoothly cover ordinary space. This means that we no longer consider the 
configuration - or coordinate - space description as fundamental; indeed the spatial points $\{\bold{x}\}$ no longer have any meaning outside of the spectra of the corresponding operator in our algebra (such as $\hat{x}$). In other words, the $\{\bold{x}\}$ are merely a collection of eigenvalues, and are no longer to be used in the classical sense to define intervals $\Delta \bold{x} = \bold{x_2} - \bold{x_1}$. 

In NCG, the torsion-gravity action (\ref{fullacn}) is replaced by the NCG torsion-gravity action of the type
\begin{equation}
S_{NCG} =\frac{1}{L_{pl}^2} \intbar_{geom}d\hat s^2\quad  + \quad \frac{1}{L^2} \intbar_{matter} d\hat s^2
\label{ncga}
\end{equation}
(where we have assumed $L_1=L_2$). In the Riemannean limit, when the coordinates become commuting, this action reduces to the action (\ref{fullacn}). We call action (\ref{ncga}) an atom of space-time-matter (STM). Thus, we do not make any distinction between matter (i.e. the elementary particle/closed string) and the non-local space-time geometry it `produces'. The STM atom is a primordial unit of structure, as if to say that the matter unit carries its own space-time geometry and is a universe in itself. The distinction between matter and 
space-time geometry is only an emergent concept. The full Hilbert space is populated by a large number of STM atoms, each carrying its own NCG space-time. This is depicted as Level 0 in Fig. 1. Various atoms interact via entanglement - say $\ket{1a}$ and $\ket{1b}$ are two states of the first atom, and similarly $\ket{2a}$ and $\ket{2b}$ for the second atom, then the entangled state $\ket{1a}\ket{2a} + \ket{1b}\ket{2b}$ represents interaction, including the gravitational interaction of the two space-time geometries. We will see below that when a very large number of STM atoms are entangled, that leads to the emergence of classical space-time.

The non-commutative action in principle implies a set of operator field equations which relate the closed string to its curvature-torsion. In NCG, according to the Tomita-Takesaki theory, there is a one-parameter group of inner automorphisms of the algebra ${\cal A}$ of the coordinates - this serves as a `god-given' [as Connes puts it] time parameter with respect to which non-commutative spaces evolve. This Connes time $\tau$ has no analog in the commutative case, and we employ it to describe evolution. An STM atom evolves geodesically in Connes time. The evolution is non-linear;  it is non-unitary (because the asymmetric metric is not necessarily self-adjoint), yet it is norm-preserving because the evolution is geodesic. These are precisely the features required for spontaneous collapse in models of objective wave vector reduction. But unlike the ad hoc noise introduced in collapse models, here the desired GRW/CSL type evolution is natural to an STM atom in non-commutative geometry. As we explain below, from here the physics of spontaneous collapse is emergent at Level I. 

In the spirit of the theory of trace dynamics, one can use the above non-commutative action to define canonical configuration and momentum dynamical variables, and from there a trace Hamiltonian. It is expected that the theory admits a conserved Adler-Millard charge $\tilde{C}$, this being the sum of the $[q,p]$ commutators over all the bosonic degrees of freedom, minus the sum over anti-commutators $\{q,p\}$ of the fermionic degrees of freedom.  Like in ordinary classical dynamics, Hamilton's equations of motion can be worked out as well. In this regard, the NCG of an STM atom resembles the matrix mechanics of trace dynamics \cite{Adler:04}, but with the space-time manifold replaced by the non-local non-commutative space of NCG.Thus the rules of calculus in this matrix dynamics are those of NCG.

Assuming that one is observing the dynamics of the STM atoms over length scales much larger than Planck length, one constructs the statistical thermodynamics of a large number of STM atoms, taking us to Level I. An entropy function and a partition function are constructed, and maximising the entropy yields the theory at thermodynamic equilibrium. The thermal bath with which the STM atoms are assumed to be in equilibrium requires an absolute time  parameter to be specified, and we assume that to be Connes time. Since there is no classical space-time at this level, we do not consider this as a violation of Lorentz invariance. By the time Lorentz invariance emerges at Level III, non-commutativity is lost, and hence the Connes time is lost too.

The full Hamiltonian is the sum of Hamiltonians of individual STM atoms, and 
at thermodynamical equilibrium, the conserved Adler-Millard charge is equipartitioned over all the degrees of freedom, with the constant equipartitioned value per each degree of freedom assumed to be equal to the Planck constant $\hbar$. Thus in $\hbar$ we have the third fundamental constant of the theory, after Planck length and speed of light; then Newton's gravitational constant $G$ is defined as $G \equiv L_{pl}^2 \; c^3 / \hbar$. As in trace dynamics, the canonical thermal averages of the dynamical variables obey the standard commutation relations of quantum field theory, and they also obey the Heisenberg equations of motion. We have arrived at Level I.

In the functional Schr\"{o}dinger picture, at equilibrium, one gets a Wheeler-DeWitt like equation, for the state $\Psi_i$ of the gravitational (i.e. asymmetric metric) and matter degrees of freedom of the $i$-{\it th} STM atom:
\begin{equation} 
i\hbar \frac{\delta \Psi_i}{ \delta \tau} = H_i \; \Psi_i 
\end{equation}
One is still in the same Hilbert space as that of Level 0, but we now have a coarse-grained view of this space. We are seeing the approximate equations satisfied by the STM atoms, in thermodynamic equilibrium, and these are the equations of quantum gravity. Quantum gravity is an emergent phenomenon, arising from an underlying non-commutative classical dynamics operating in the Hilbert space at the Planck scale. As before, one can make entangled states from solutions of the Wheeler-deWitt like equations for different STM atoms.

As in trace dynamics \cite{Adler:04} statistical fluctuations around equilibrium play an extremely significant role in this theory. These fluctuations prevent the equipartition of the 
Adler-Millard charge, and result in departures from quantum theory and from quantum general relativity.  The impact of fluctuations on the averaging of $\tilde{C}$ can be represented as stochastic corrections to its ensemble average. This in turn results in stochastic modification of the emergent Heisenberg equations of motion. Equivalently, the
 Wheeler-deWitt like functional Schr\"{o}dinger equation gets endowed with stochastic modifications, because the fluctuations come from corrections to the Adler-Millard charge.
  While the STM Hamiltonians for different atoms decouple from each other, the fluctuations couple different STM atoms, because the corrections to $\tilde{C}$ depend on all the atoms, and not necessarily in a separable way. Thus the stochastic 
 Wheeler-deWitt like equation takes the non-linear form
 \begin{equation}
 i\hbar \frac{\delta \Psi (q_1,q_2,...)} {\delta \tau}  = \left(\left[ \sum_{i} H_i(q_i)\right] + {\mathcal K} (\Psi, q_1,q_2,...) \right) \Psi(q_1,q_2,...)
 \label{stocwd}
 \end{equation}
 where ${\mathcal K}$ represents  stochastic corrections (which naturally includes an anti-self-adjoint part and depends on the state $\Psi$) to the quantum gravity equation. This non-unitary and non-linear stochasticity is a reflection of the underlying theory at Level 0. It results in spontaneous localisation, taking us from Level I to III.
 
 Spontaneous localisation has been proposed earlier as a mechanism to explain the quantum-to-classical transition in quantum physics, and to provide a falsifiable solution to the quantum measurement problem \cite{Ghirardi:86, Ghirardi2:90}. However, we see here that the idea of spontaneous collapse has  much greater significance than the measurement problem. A relativistic generalisation of spontaneous localisation, as in (\ref{stocwd}), suggests that space-time in itself arises from this collapse of the wave function, and the solution of the measurement problem is merely a special case of this phenomenon \cite{Singh:2018, Singh:2019}. The process operates as follows.

Consider a state $\Psi$ of $N$ STM atoms in the Hilbert space at Level I, labeled by the space-time operator coordinates $\hat{x}^{i}_n$ of the various atoms. Thus 
$\Psi= \Psi (\tau, {\hat x}^{i}_1, \hat{x}^{i}_2, ...., \hat{x}^{i}_N)$. 
At random $\tau$ time, the matter part of the $n$-th atom undergoes spontaneous localisation to some random eigenvalue $x^{i}_n$ of its operator coordinate.  The (non-matter) curvature part of the STM atom stays uncollapsed. Macroscopic states are those which involve entanglement of many atoms - these undergo very  rapid spontaneous collapse, and are responsible for the emergence of classical Lorentzian space-time.  The classical field equations of Level III arise as follows. Upon spontaneous collapse, each atom obeys the field equations of commutative torsion gravity as given above in Section II. The average space-time produced by many STM atoms is the average of the individual space-times of the various STM atoms: the metric, connection, and curvature are all averaged, e.g. the curvature scalar $R$ of the emergent classical space-time is the average of the curvature scalars $R_n$ of the individual atoms:
$R = <\{R_n\}>$. With this assumption, the field equations of the emergent universe are the same as the equations  of classical torsion gravity, with the matter source being given by energy-momentum coming from Eqn. (\ref{6a}), the sum over $n$ now representing the sum over many particles. When spin can be neglected, these reduce to the field equations of classical general relativity.
Uncollapsed objects stay on Level I -- this means that fluctuations are not significant for them. For such one or more STM atom(s), dynamics is described by the Wheeler-deWitt like equation (without fluctuations). There is no background space-time, nor a gravitational interaction, but entanglement is possible.

Level II describes quantum field theory on a classical curved background, and in particular on a Minkowski background.  Strictly speaking, the quantum fields of Level II should be described at Level I, because as we have argued earlier, one cannot fundamentally have matter as quantum and simultaneously space-time as classical. Except by making appropriate approximations and assumptions at Level I. These approximations are as follows: a) The curvature produced by the STM atoms at Level I is neglected. b) In the resulting operator space-time, the anti-symmetric part of the metric is suppressed, so that one is left only with an operator  Minkowski space-time. Since every STM atom now has an associated Minkowski space-time, it is assumed that together these   individual Minkowski space-times are equivalent to a global Minkowski operator space-time. Thus we have quantum fields on an operator  Minkowski 
space-time. In analogy with the Stueckelberg many-particle approach to relativistic quantum field theory, we equivalently describe this system as relativistic quantum mechanics on an operator space-time \cite{Singh:2018, Singh:2019}. 

Suppressing the operator nature of coordinate time facilitates the transition from Level I to Level II. At the same time, the operator nature of time is responsible for the novel phenomenon of quantum interference in/of time, for which we have argued that there is experimental evidence \cite{Singh:2019, Horwitz:2006, L2005}. The operator nature of spatial coordinates at Level I is interchanged for the operator nature of the canonical variables of quantum fields at Level II.This is the justification for the conventional quantisation conditions when one transits from Level III to Level II. When the operator nature of time is suppressed, Connes time is interchanged for coordinate time, in going from Level I to Level II.
The transition from level II to Level III takes place through  a GRW-type non-relativisitic spontaneous localisation \cite{Singh:2018}. In particular such localisation solves the quantum measurement problem. 

In the absence of non-gravitational interactions, spontaneous localisation from Level I to III results in the formation of black holes. Being away from thermodynamic equilibrium, these are states of [relatively] low entropy. Via Hawking evaporation, a black hole attempts to return back to the maximum entropy thermodynamic equilibrium state of Level I. In our picture, the formation of classical black holes and the emergence of the space-time manifold and of classical general relativity are far from equilibrium processes. Eventually, over astronomical periods, Hawking radiation takes the universe back to the quantum equilibrium of Level I. And the inter-play between equilibrium and fluctuations continues  in Connes time.

\bigskip
\newpage

\centerline{\bf REFERENCES}

\bibliography{biblioqmtstorsion}

\end{document}